\documentclass[sigconf]{acmart}
\usepackage{booktabs} 
\usepackage{amsmath,url,graphicx,amssymb}
\usepackage{amsfonts}
\usepackage{ctable}
\usepackage{multirow}
\usepackage{algorithm}
\usepackage{algpseudocode}
\usepackage{pifont}
\usepackage{color}
\usepackage{subfigure}
\usepackage{enumitem}
\usepackage{sidecap}

\newcommand{\rom}[1]{\uppercase\expandafter{\romannumeral #1\relax}}

\newcommand{\mylistbegin}{
  \begin{list}{$\bullet$}
   {
     \setlength{\itemsep}{-2pt}
     \setlength{\leftmargin}{1em}
     \setlength{\labelwidth}{1em}
     \setlength{\labelsep}{0.5em} } }
\newcommand{\mylistend}{
   \end{list}  }

\newcommand{\eg}{\textit{e.g.}}

\newcommand{\ie}{\textit{i.e.}}
\newcommand{\etc}{\textit{etc}}

\newcommand{\wrt}{\textit{w.r.t.~}}

\newcommand{\header}[1]{{\vspace{+1mm}\flushleft \textbf{#1}}}

\newtheorem{thm:def}{Definition}

\setcopyright{rightsretained}
\acmDOI{}
\acmISBN{}

\acmConference[KDD'19 Demo]{Proceedings of the 25th ACM SIGKDD International Conference on Knowledge Discovery and Data Mining}{Anchorage, AK}{August 3-8, 2019} 
\acmYear{2019}
\copyrightyear{2019}

\acmPrice{15.00}

\begin{document}
\title{CubeNet: Multi-Facet Hierarchical Heterogeneous Network Construction, Analysis, and Mining}
\author{Carl Yang$^*$, Dai Teng$^*$, Siyang Liu$^*$, Sayantani Basu$^*$, Jieyu Zhang$^*$, Jiaming Shen$^*$, Chao Zhang$^*$, Jingbo Shang$^*$, Lance Kaplan$^\#$, Timothy Harratty$^\#$, Jiawei Han$^*$}
       \affiliation{
       \institution{$^*$University of Illinois at Urbana Champaign, 201 N Goodwin Ave, Urbana, IL 61801, USA}
       \institution{$^\#$US Army Research Laboratory, 2800 Powder Mill Rd, Adelphi, MD 20783, USA}
       \institution{$^*$\{jiyang3, daiteng2, sliu134, basu9, jieyuz2, js2, czhang82, shang7, hanj\}@illinois.edu, \\$^\#$\{lance.m.kaplan.civ, timpthy.p.hanratty.civ\}@mail.mil}}

\setlength{\floatsep}{4pt plus 4pt minus 1pt}
\setlength{\textfloatsep}{4pt plus 2pt minus 2pt}
\setlength{\intextsep}{4pt plus 2pt minus 2pt}

\setlength{\dbltextfloatsep}{3pt plus 2pt minus 1pt}
\setlength{\dblfloatsep}{3pt plus 2pt minus 1pt}
\setlength{\abovecaptionskip}{3pt}
\setlength{\belowcaptionskip}{2pt}
\setlength{\abovedisplayskip}{2pt plus 1pt minus 1pt}
\setlength{\belowdisplayskip}{2pt plus 1pt minus 1pt}
\renewcommand\footnotetextcopyrightpermission[1]{}
\pagenumbering{gobble}
\renewcommand{\shortauthors}{Carl Yang \textit{et al.}}
\settopmatter{printacmref=false, printfolios=false}

\begin{abstract}
Due to the ever-increasing size of data, construction, analysis and mining of universal massive networks are becoming forbidden and meaningless. 
In this work, we outline a novel framework called CubeNet, which systematically constructs and organizes real-world networks into different but correlated semantic cells, to support various downstream network analysis and mining tasks with better flexibility, deeper insights and higher efficiency.
Particular, we promote our recent research on text and network mining with novel concepts and techniques to (1) construct four real-world large-scale multi-facet hierarchical heterogeneous networks; (2) enable insightful OLAP-style network analysis; (3) facilitate localized and contextual network mining.
Although some functions have been covered individually in our previous work, a systematic and efficient realization of an organic system has not been studied, while some functions are still our on-going research tasks. By integrating them, CubeNet may not only showcase the utility of our recent research, but also inspire and stimulate future research on effective, insightful and scalable knowledge discovery under this novel framework.
\end{abstract}

\maketitle
\vspace{-10pt}
{\small \textbf{ACM Reference Format:}\\
Carl Yang \textit{et al}. 2019. CubeNet: Multi-Facet Hierarchical Heterogeneous Network Construction, Analysis, and Mining. In \textit{Proceedings of KDD 2019 (Demo), August 3-8, 2019, Anchorage, AK}, 2 pages.}
\section{Introduction}
\label{sec:intro}
\vspace{-2pt}
\header{Research Project, Goals and Partners.}
Real-world networks nowadays are becoming very large (\eg, DBLP publication network with 4 millions of paper nodes, Facebook social network with 2 billions of user nodes, \etc). 
It is hard and wasteful, if not impossible, for various algorithms to scale up to the sheer sizes of networks, and many network analytical measures and mining tasks become meaningless on the massive universal networks. 

In this work, we demonstrate the ability and value of constructing and organizing massive networks \wrt~an underlying data cube structure. 
Firstly, based on metadata and textual contexts, we propose to automatically construct multi-facet hierarchical data cube structures for semantic-aware organization of massive networks. 
Next, we enable various cube-based OLAP-style network analysis including contrastive network summarization, cell-based semantic backtracking and multi-granularity structure exploration.
Finally, we develop novel concepts and techniques for flexible and insightful network mining including contextual pattern mining, query-specific network localization and conditional proximity search.


\header{Fit with the KDD ecosystem}
This project will be attractive to the large audience interested in network mining, network science, text mining, big data management and various downstream applications. While researchers and practitioners familiar with network data mining and data management may get mostly benefited, we believe it will also provide valuable insight towards the new era of network science and inspire general audience and even newcomers to KDD. Particularly, the series of state-of-the-art techniques we develop and showcase brings new light to how massive network data could be organized and utilized in the future. The project and leveraged techniques are all fully open-sourced, so the audience can build their own CubeNet following our approaches in an effortless manner. 

\vspace{-3pt}
\section{Main Innovations}
\label{sec:model}
\begin{figure*}[t]
\vspace{-10pt}
\centering
\includegraphics[width=0.9\textwidth]{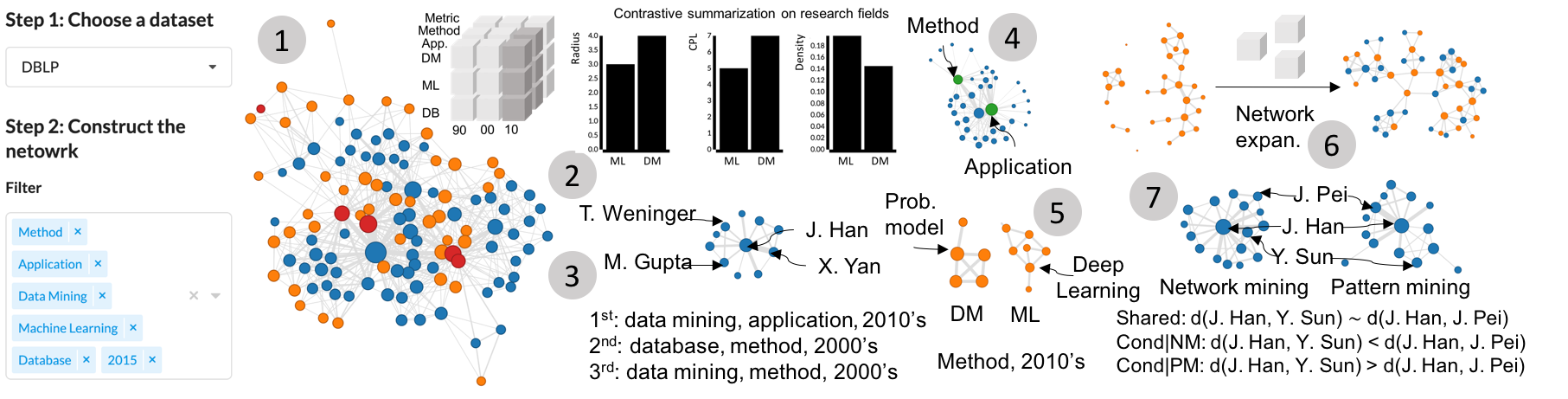}
\vspace{-2pt}
\caption{\textbf{An overview of the proposed CubeNet system with examples illustrating the novel functions we developed.}}
\label{fig:toy}
 \end{figure*}

Figure \ref{fig:toy} gives an overview (Ex.~1) of the proposed CubeNet system, where a large multi-facet heterogeneous network is organized \wrt~a topic-objective-year data cube structure.
 
 \vspace{-5pt}
\subsection{CubeNet Construction}
\vspace{-3pt}
\header{Heterogeneous network enrichment.}
Without clear semantics, real-world networks are less informative. For more insightful data analysis and mining, we enrich the heterogeneity of networks by incorporating nodes from massive free texts. 
In this system, we leverage our recent research on text mining, \ie, AutoPhrase \cite{shang2018automated} for phrasal node extraction and AutoNER \cite{shang2018learning} for typed link generation.

\vspace{-3pt}
\header{Multi-facet taxonomy generation.}
In this system, we leverage both existing metadata and our recent research on automatic taxonomy generation, \ie, TaxoGen \cite{zhang2018taxogen}, to create multiple taxonomies for each network, essentially leading to a data cube structure \cite{tao2016multi}. 

\vspace{-3pt}
\header{Weakly-supervised network organization.}
To organize networks into the data cube structure, \ie, allocate nodes to proper cells, we leverage our recent research on heterogeneous network classification based on AutoPath \cite{yang2018similarity}, which assigns similar labels in the taxonomy to nodes close in the network based on small sets of nodes weakly labeled via surface name matching.

\vspace{-3pt}
\subsection{CubeNet Analysis}
\vspace{-2pt}
\header{Contrastive network summarization.}
Various traditional network statistical measures such as clustering coefficient, character path length and triangle count become hard to compute and meaningless in massive universal networks. However, in CubeNet where each cell holds a relatively small network, the structures can be efficiently summarized and contrasted across relevant cells by aggregating network statistics, which provides insight into network evolution along different semantic dimensions (Ex.~2 in Figure \ref{fig:toy}).

\vspace{-3pt}
\header{Cell-based semantic backtracking.}
While text cube supports the retrieval of most relevant cells \wrt~unary queries, CubeNet further allows semantic backtracking \wrt~network queries, such as pairs of nodes and small sub-networks. The idea is to combine the graph coverage \cite{shen2018entity} and top-$k$ cell search \cite{ding2011efficient} algorithms to find the $k$ cells that mostly cover the network query from all cells with an optimized search order (Ex.~3 in Figure \ref{fig:toy}).

\vspace{-3pt}
\header{Multi-granularity structure exploration.}
By allocating nodes into hierarchically organized cells, CubeNet supports network roll-up and drill-down, which essentially merges nodes and edges into super-nodes and super-edges (or the other way around), to allow the exploration of network structures in preferred granularities. To make the process efficient, we implement the techniques developed in our previous research on graphcube \cite{zhao2011graph} (Ex.~4 in Figure \ref{fig:toy}).

\vspace{-5pt}
\subsection{CubeNet Mining}
\vspace{-3pt}
\header{Contextual network pattern mining.}
Traditional graph pattern mining does not consider the contexts of networks. To find more semantically related patterns, we extend \cite{yan2003closegraph} to CubeNet by computing a mixture score of popularity, integrity and distinctiveness. Popularity is computed as the normalized frequency, integrity is the ratio of frequency between the pattern and its corresponding close pattern, and distinctiveness is the ratio between the frequency in a cell and the average of all cells. They are combined using customized weights to highlight user preference (Ex.~5 in Figure \ref{fig:toy}).

\vspace{-3pt}
\header{Query-specific network localization.}
Given data mining queries over particular sets of nodes, computation over the universal massive network is wasteful and hard to handle. Since CubeNet organizes networks by grouping semantically relevant nodes, it is possible to find a set of cells that mostly cover the queried nodes. To this end, we apply our on-going research on query-specific network construction, which leverages a light reinforcement learning algorithm to find the optimal combination of cells, from which a relevant and complete network can be constructed to support downstream data mining on queried set of nodes (Ex.~6 in Figure \ref{fig:toy}).

\vspace{-3pt}
\header{Conditional proximity search.}
Recent research on network embedding often does not easily scale to massive networks and fail to model node proximity under different semantic conditions \cite{yang2018meta}. To deal with both challenges, we apply our on-going research on the co-embedding of network nodes and cube cells, which jointly learns node embedding in each sub-network and a set of embedding transformation functions that align relevant sub-networks. The embedding of sub-networks thus facilitates proximity search conditioned on cell semantics, while the alignment functions enable proximity transfer among similar cells (Ex.~7 in Figure \ref{fig:toy}).

\section{Demonstration}
\label{sec:demo}
The current CubeNet system indexes 4 real-world large-scale heterogeneous networks: (1) a DBLP network with 2M nodes of authors, topical phrases, venues, years and 0.4B links; (2) a Yelp network with 0.8M nodes of businesses, opinion phrases, locations, stars and 0.2B links; (3) a PubMed network with 0.2M nodes of genes, proteins, diseases, chemical compounds, species and 0.1B links; (4) a FreeBase network with 16M nodes of persons, locations, books, movies, \etc.~and 77M links. We further generate taxonomies of 11K topics for DBLP, 70K categories for Yelp, 197K diseases for PubMed, and 800K types for FreeBase, together with other trivial taxonomies like publication years and rating scores from metadata. A toy system is available at https://github.com/yangji9181/CubeNet, which currently supports main functionalities on DBLP and Yelp, while a full version will be gradually rolled out before demonstration. 
\vspace{-5pt}
\section*{Acknowledgement}
Research was sponsored in part by U.S. Army Research Lab. under Cooperative Agreement No. W911NF-09-2-0053 (NSCTA), DARPA under Agreement No. W911NF-17-C-0099, National Science Foundation IIS 16-18481, IIS 17-04532, and IIS-17-41317, DTRA HDTRA11810026, and grant 1U54GM114838 awarded by NIGMS through funds provided by the trans-NIH Big Data to Knowledge (BD2K) initiative (www.bd2k.nih.gov).
\bibliographystyle{ACM-Reference-Format}
\bibliography{carlyang} 

\end{document}